\documentstyle[12pt]{article}
\def\g{\gamma}
\def\p{\psi}

\def\l{\lambda}

\def\be{\begin{equation}}
\def\ee{\end{equation}}
\def\arr{\begin{array}{rll}}
\def\ea{\end{array}}
\def\bea{\begin{eqnarray}}
\def\eea{\end{eqnarray}}

\begin{document}
\renewcommand{\thefootnote}{\fnsymbol{footnote}}
\begin{titlepage}
\noindent

\vskip 3.0cm

\begin{center}

{\Large\bf Remarks on N=4 Superconformal }\\

\bigskip

{\Large\bf Extension of the Calogero Model }\\

\bigskip

\vskip 1cm

{\large Anton V. Galajinsky}\footnote{galajin@mph.phtd.tpu.edu.ru, galajins@lnf.inf.it}

\vskip 0.5cm

{\it Laboratory of Mathematical Physics, Tomsk Polytechnic University, \\
634050 Tomsk, Lenin Ave. 30, Russian Federation}\\

\vskip 0.2cm
{\it and }\\
\vskip 0.2cm
{\it INFN--Laboratori Nazionali di Frascati, C.P. 13,
00044 Frascati, Italy}\\

\vskip 0.2cm

\end{center}
\vskip 1cm

\begin{abstract}
Recently it was conjectured by Gibbons and Townsend that the large $n$ limit of
an $N=4$ superconformal extension of the $n$--particle Calogero model might provide a
microscopic description of the extreme Reissner--Nordstr\"om black hole near the horizon.
In this paper a possibility to construct an $SU(1,1|2)$ invariant extension
of the Calogero model is considered. We treat in detail the two--particle case and
comment on some peculiarities intrinsic to $n>2$ generalizations.
\end{abstract}

\vspace{0.5cm}


\end{titlepage}
\renewcommand{\thefootnote}{\arabic{footnote}}
\setcounter{footnote}0
\noindent

The Calogero model~\cite{cal} has many interesting physical applications.
It provides a non\-tri\-vial example of $n$--particle integrable
system in one dimension~(see Ref.~\cite{per} for a review). Calogero-type models are
related to semisimple Lie algebras~\cite{per} and reveal an intriguing connection with
Hamiltonian reductions of the $2d$ Yang--Mills theory~\cite{gor}. Higher dimensional
generalizations appear in the context of fractional statistics~\cite{frac}
and higher spin gauge theories~\cite{vas}. In quantum area the one--dimensional system
turns out to be completely solvable (see e.g. Ref.~\cite{vas} and references therein).

Quite recently the $n$--particle Calogero model has been brought to focus again, this time
in the context of black hole physics~\cite{gib}. It has been known for a long time~\cite{reg}
that the $n$--particle Calogero model with the Hamiltonian
\be\label{calogero}
H=\frac{1}{2}{\sum}_{i=1}^n~ p_i^2  +\frac{1}{2} {\sum}_{i<j}~ \frac{g^2}{{(x_i-x_j)}^2},
\ee
where $g$ is a dimensionless coupling constant, exhibits the conformal invariance. The charges
\be\label{charges}
D=tH-\frac{1}{2} {\sum}_{i=1}^n~ x_i p_i, \quad K=t^2 H -t{\sum}_{i=1}^n~ x_i p_i +\frac{1}{2}{\sum}_{i=1}^n x_i x_i,
\ee
are conserved and together with the Hamiltonian~(\ref{calogero}) form the $so(1,2)$ algebra
\be
\{H,D \}=H, \quad \{H,K \}=2D, \quad \{D,K \}=K,
\ee
which is the conformal algebra in one dimension. The conjecture~\cite{gib} that the $n$--particle
Calogero model might be relevant for a microscopic description of the extreme Reissner--Nordstr\"om black hole,
at least near the horizon, originated from the fact that the near horizon geometry for this case
has the $SU(1,1|2)$ isometry group, the bosonic subgroup being $SO(1,2)$. Taking into account that
the extreme Reissner--Nordstr\"om black hole can be viewed as the configuration of
four intersecting supergravity $D3$--branes wrapped on $T^6$~\cite{kleb} and assuming that each of the supergravity
$D3$--branes can be interpreted as a large number of coinciding microscopic $D3$--branes~\cite{gib},
one comes to the conclusion that there must be an $SU(1,1|2)$ invariant mechanics which governs the
fluctuation of the branes in the region of the intersection~\cite{gib}. Being conformally invariant, the
$n$--particle Calogero system provides a possible candidate\footnote{In the context of black hole
physics the large $n$ limit is to follow~\cite{gib}.} for the bosonic part of such an
$N=4$ superconformal mechanics.

In this brief note we discuss a possibility to construct an $N=4$ superconformal extension of the
Calogero model. We treat in detail the two--particle case and comment on some peculiarities intrinsic
to $n>2$ generalizations. Throughout the paper we choose to work in the Hamiltonian formalism.

As the first step one has to decide on the number of fermions to be assigned to each particle.
Because the system must accommodate an $N=4$ supersymmetry and since in the one-particle limit
one expects to reproduce the $N=4$ superconformal mechanics~\cite{ikl} it seems
na\-tu\-ral to append a pair of complex fermions ${(\psi_i)}^{*}=\bar\psi_i$, $i=1,2$,
obeying the bracket $\{\psi_i,\bar\psi_j \}=-i\delta_{ij}$, to each bosonic canonical pair
$(x_1,p_1)$ and $(x_2,p_2)$.
Let us mark the fermionic degrees of
freedom corresponding to each particle by the superscripts $(1)$ and $(2)$,
respectively\footnote{In order to make the formulae more readable we loosely put the index $i$
on both $\p$ and $\bar\p$ down. A more proper notation would be
${(\psi^i)}^{*}=\bar\psi_i$,  $\{\psi^i,\bar\psi_j \}=-i{\delta^i}_j$.}.
Apart from the conformal generators $H,D,K$, the $su(1,1|2)$ superalgebra includes two complex
supersymmetry charges $G_1,G_2$ (their conjugates will be denoted by $\bar G_1, \bar G_2$), the superconformal partners
$S_1,S_2,\bar S_1,\bar S_2$, and the $R$--symmetry $su(2)$ generators $J_{+},J_{-},J_3$
(in the basis chosen the commutation relations of the $su(1,1|2)$ superalgebra are given in Appendix).
As the bracket $\{G_i,\bar G_j \}=-2iH \delta_{ij}$, $i,j=1,2$, makes part of the
superalgebra, the contributions to $G_i$, $\bar G_i$ which are linear in the fermions should be adjusted so as to
produce the Hamiltonian~(\ref{calogero}) (with $n=2$ for the case at hand). This is typical of supersymmetric (quantum)
mechanics. Another important point to notice is a representation
of the $su(2)$ subalgebra. This is not unique and normally one chooses a "direct sum"
representation (in this respect see also Refs.~\cite{wyl},\cite{ghosh})
\bea\label{su2}
&&
J_{+}=-i\psi^{(1)}_1 {\bar\psi}^{(1)}_2 -i \psi^{(2)}_1 {\bar\psi}^{(2)}_2, \quad
J_{-}=i\psi^{(1)}_2 {\bar\psi}^{(1)}_1 +i \psi^{(2)}_2 {\bar\psi}^{(2)}_1,
\nonumber\\[2pt]
&& J_3=\textstyle{\frac{1}{2}}(\psi^{(1)}_1 {\bar\psi}^{(1)}_1 -\psi^{(1)}_2 {\bar\psi}^{(1)}_2+
\psi^{(2)}_1 {\bar\psi}^{(2)}_1 -\psi^{(2)}_2 {\bar\psi}^{(2)}_2),
\eea
which implies that the fermions associated with each particle form a separate $SU(2)$ doublet.
Because the supersymmetry generators transform under the action of $SU(2)$
(the brackets are given in the basis chosen)
\bea\label{ferm}
&&
\{G_1,J_{+} \}=0, \quad \{G_1,J_{-} \}=\bar G_2, \quad \{G_1,J_3 \}=\textstyle{\frac{i}{2}} G_1,
\nonumber\\[2pt]
&&
\{G_2,J_{+} \}=0, \quad \{G_2,J_{-}\}=-{\bar G}_1, \quad \{G_2,J_3\}={\textstyle{\frac{i}{2}}}G_2,
\eea
the structure of the fermionic contributions in $G_1$ and $G_2$ is completely fixed by the latter relations
and the representation~(\ref{su2}). As to the bosonic coefficients in front of the fermions, the fact that
$G_1$ is nilpotent suggests a passage to the center of mass and relative coordinates
$(x_1+x_2,p_1+p_2)$, $(x_1-x_2,p_1-p_2)$. After adjusting the coefficients in a proper way one
finds the following representation for the $N=4$ supersymmetry generators
\bea
&&
G_1=\frac{1}{\sqrt{2}} (p_1-p_2) \p^{(1)}_1 +\frac{g}{x_1-x_2} {\bar\p}^{(1)}_2
+\frac{2i}{\sqrt{2}} \frac{1}{x_1-x_2} \p^{(1)}_1  \p^{(1)}_2  {\bar\p}^{(1)}_2+
\nonumber\\[2pt]
&& \qquad
+\frac{1}{\sqrt{2}} (p_1+p_2) \p^{(2)}_1
+\frac{2i}{\sqrt{2}} \frac{1}{x_1+x_2} \p^{(2)}_1  \p^{(2)}_2  {\bar\p}^{(2)}_2,
\nonumber\\[2pt]
&&
G_2=-\frac{1}{\sqrt{2}} (p_1-p_2) {\bar\p}^{(1)}_2 +\frac{g}{x_1-x_2} \p^{(1)}_1
+\frac{2i}{\sqrt{2}} \frac{1}{x_1-x_2} \p^{(1)}_1  {\bar\p}^{(1)}_1  {\bar\p}^{(1)}_2-
\nonumber\\[2pt]
&& \qquad
-\frac{1}{\sqrt{2}} (p_1+p_2) {\bar\p}^{(2)}_2
+\frac{2i}{\sqrt{2}} \frac{1}{x_1+x_2} \p^{(2)}_1 {\bar\p}^{(2)}_1  {\bar\p}^{(2)}_2,
\eea
which yield the Hamiltonian
\bea\label{ham}
&&
H=\frac{1}{2}p_1^2 + \frac{1}{2}p_2^2 +\frac{g^2}{{2(x_1-x_2)}^2}+
\frac{2ig}{\sqrt{2} {(x_1-x_2)}^2}~ \p^{(1)}_1 \p^{(1)}_2+
\frac{2ig}{\sqrt{2} {(x_1-x_2)}^2}~ {\bar\p}^{(1)}_1 {\bar\p}^{(1)}_2+
\nonumber\\[2pt]
&& \qquad
+\frac{2}{{(x_1-x_2)}^2}~\p^{(1)}_1 {\bar\p}^{(1)}_1  \p^{(1)}_2 {\bar\p}^{(1)}_2+
\frac{2}{{(x_1+x_2)}^2}~\p^{(2)}_1 {\bar\p}^{(2)}_1  \p^{(2)}_2 {\bar\p}^{(2)}_2.
\eea
Because the pairs $G_1,\bar G_1$ and $G_2,\bar G_2$ mutually commute, the conservation in time
of these generators is guaranteed by Jacobi identities.

To construct a representation for
the superconformal generators $S_1,\bar S_1, S_2,\bar S_2$ one appeals to the
$su(1,1|2)$ superalgebra (see Appendix). It suffices to calculate the Poisson bracket of the
supersymmetry charges with the generator of special conformal transformations $K$
(see Eq.~(\ref{charges}) above with $H$ taken from Eq.~(\ref{ham})) which gives
\bea
&&
S_1=tG_1-{\textstyle\frac{1}{\sqrt{2}}} (x_1-x_2) \p^{(1)}_1
-{\textstyle\frac{1}{\sqrt{2}}} (x_1+x_2) \p^{(2)}_1,
\nonumber\\[2pt]
&&
S_2=tG_2+{\textstyle\frac{1}{\sqrt{2}}} (x_1-x_2) {\bar\p}^{(1)}_2
+{\textstyle\frac{1}{\sqrt{2}}} (x_1+x_2) {\bar\p}^{(2)}_2.
\eea
It is straightforward to check the the full algebra is closed (the commutation relations are gathered
in Appendix) and we conclude that the Hamiltonian~(\ref{ham}) governs the dynamics of an
$SU(1,1|2)$ invariant extension of the two--particle Calogero model. Notice that in the one--particle limit
(i.e. setting $x_2=0$, $p_2=0$, $\p^{(2)}_i=0$)
the system~(\ref{ham}) reproduces the $N=4$ superconformal mechanics of Ref.~\cite{ikl}
(with the parameters $c=0$, $f=1$; see~\cite{ikl} for more details).
A few comments are in order.\\
i) Let us compare our result with the previous attempts~\cite{wyl},\cite{ghosh} to construct an
$N=4$ superconformal extension of the Calogero model. According to the analysis of
Ref.~\cite{wyl} for generic values of the coupling constant $g$ and for $n>2$ there is no way to build
an $SU(1,1|2)$ invariant extension, while for $n=2$ this can be done only for a particular value of the
coupling constant (see Ref.~\cite{wyl} for more details). As we have seen above, taking a more general ansatz
for the supersymmetry generators one can weaken the restriction on the coupling constant and construct a two--particle
model for generic values of the coupling constant\footnote{A passage to the center of mass and relative coordinates in the
Hamiltonian of the two--particle Calogero model brings it to the sum of a free particle and the conformal
mechanics~\cite{aff}, both admitting an $N=4$ superconformal extension. Thus, the existence of an $SU(1,1|2)$
invariant extension of the two--particle Calogero model for an arbitrary value of the coupling constant
might be anticipated on this general ground. We thank Sergey Krivonos for pointing this out to us.}.
In Ref.~\cite{ghosh} an $SU(1,1|2)$--invariant model in two dimensions has been constructed without any restrictions
on the coupling constant or the number of particles. Since a naive dimensional reduction to one dimension does not
preserve all four supersymmetries it seems interesting to search for a more sophisticated reduction.\\
ii) Although we did not try to systematically extend our analysis to the $n>2$ case, a preliminary consideration
shows that this may require a modification of the $su(2)$ representation~(\ref{su2}).
One such possibility is to mix fermions in the $su(2)$--generators so that the fermions belonging to
different particles share the same $SU(2)$ spinor representation. For example, for the
three--particle case the generators
\bea\label{sumod}
&&
J_{+}=\frac{1}{\g} \left[-i\psi^{(1)}_1 {\bar\psi}^{(1)}_2 -i \psi^{(2)}_1 {\bar\psi}^{(2)}_2
-i\psi^{(3)}_1 {\bar\psi}^{(3)}_2+\l(\psi^{(1)}_1 {\bar\psi}^{(2)}_2+
\psi^{(2)}_1 {\bar\psi}^{(1)}_2+\psi^{(1)}_1 {\bar\psi}^{(3)}_2+\psi^{(3)}_1 {\bar\psi}^{(1)}_2+
\right.
\nonumber\\[2pt]
&& \qquad \qquad
\left.
+ \psi^{(3)}_1 {\bar\psi}^{(2)}_2+\psi^{(2)}_1 {\bar\psi}^{(3)}_2)\right],
\nonumber\\[2pt]
&&
J_{-}=\frac{1}{\bar\g} \left[i\psi^{(1)}_2 {\bar\psi}^{(1)}_1 +i \psi^{(2)}_2 {\bar\psi}^{(2)}_1
+i\psi^{(3)}_2 {\bar\psi}^{(3)}_1+{\bar\l}(\psi^{(2)}_2 {\bar\psi}^{(1)}_1+
\psi^{(1)}_2 {\bar\psi}^{(2)}_1+\psi^{(3)}_2 {\bar\psi}^{(1)}_1+\psi^{(1)}_2 {\bar\psi}^{(3)}_1+
\right.
\nonumber\\[2pt]
&& \qquad \qquad
\left.
+ \psi^{(2)}_2 {\bar\psi}^{(3)}_1+\psi^{(3)}_2 {\bar\psi}^{(2)}_1)\right],
\nonumber\\[2pt]
&&
J_3=\textstyle{\frac{1}{2}}\left[\psi^{(1)}_1 {\bar\psi}^{(1)}_1 -\psi^{(1)}_2 {\bar\psi}^{(1)}_2+
\psi^{(2)}_1 {\bar\psi}^{(2)}_1 -\psi^{(2)}_2 {\bar\psi}^{(2)}_2+
\psi^{(3)}_1 {\bar\psi}^{(3)}_1 -\psi^{(3)}_2 {\bar\psi}^{(3)}_2
\right].
\eea
form the $su(2)$ algebra, provided $\l-\bar\l-i\l\bar\l=0$, $1+2\l\bar\l=\g\bar\g$,
the simplest solution being $\l=2i$, $\g=3$.\\
iii) A representation of the $su(1,1|2)$ algebra constructed in this paper involves a central
charge which appears in the brackets of the supersymmetry generators with the superconformal ones
and is proportional to the (dimensionless) coupling constant $g$ (see Appendix).
If one assumes that $n$--particle $SU(1,1|2)$--invariant model can be constructed along similar
lines, the central term is likely to have the form $n(n-1)g$ (in this respect see also Ref.~\cite{ghosh}).
Then, in order to preserve the algebraic structure, the large $n$ limit is to be accompanied by
taking $g$ to be small with $n(n-1)g$ fixed.\\

\vspace{0.5cm}

\noindent{\bf Acknowledgements}\\

We thank Pijush K. Ghosh and Sergey Krivonos for useful discussions.
This work was partially supported by the Ministry of Education of
Russian Federation, grant E02-2.0-7, and NATO Collaborative Linkage
grant PST.CLG. 979389.

\vspace{0.5cm}
\noindent{\bf Appendix}\\

In this appendix we expose the Poisson brackets of the generators of the $su(1,1|2)$ superalgebra in
the basis chosen (vanishing brackets are omitted)
\bea
&&
\{G_1, {\bar G}_1\}=-2iH, \quad \{G_2, {\bar G}_2\}=-2iH, \quad \{G_1,J_{-}\}={\bar G}_2,
\quad \{G_1,J_3\}={\textstyle{\frac{i}{2}}}G_1,
\nonumber\\[2pt]
&&
\{{\bar G}_1, J_{+}\}=G_2, \quad
\{{\bar G}_1,J_3\}=-{\textstyle{\frac{i}{2}}}{\bar G}_1,
\quad \{G_2,J_{-}\}=-{\bar G}_1, \quad \{G_2,J_3\}={\textstyle{\frac{i}{2}}}G_2,
\nonumber\\[2pt]
&&
\{{\bar G}_2, J_{+}\}=-G_1, \quad
\{{\bar G}_2,J_3\}=-{\textstyle{\frac{i}{2}}}{\bar G}_2, \quad \{D,G_1\}=-{\textstyle{\frac{1}{2}}}G_1,
\quad \{D,G_2\}=-{\textstyle{\frac{1}{2}}}G_2,
\nonumber\\[2pt]
&&
\{D,{\bar G}_1\}=-{\textstyle{\frac{1}{2}}}{\bar G}_1,
\quad \{D,{\bar G}_2\}=-{\textstyle{\frac{1}{2}}}{\bar G}_2, \quad \{K,G_1\}=-S_1, \quad \{K,G_2\}=-S_2,
\nonumber\\[2pt]
&&
\{K,{\bar G}_1\}=-{\bar S}_1, \quad \{K,{\bar G}_2\}=-{\bar S}_2, \quad
\{G_1,S_2\}=-2iJ_{+}, \quad \{G_1,{\bar S}_1\}=-2iD+2J_3,
\nonumber\\[2pt]
&&
\{G_1,{\bar S}_2\}=-{\textstyle{\frac{i}{\sqrt{2}}}}g, \quad
\{G_2,S_1\}=2iJ_{+}, \quad \{G_2,{\bar S}_1\}={\textstyle{\frac{i}{\sqrt{2}}}}g,
\quad \{G_2,{\bar S}_2\}=-2iD+2J_3,
\nonumber\\[2pt]
&&
\{{\bar G}_1,S_1\}=-2iD-2J_3, \quad \{{\bar G}_1,S_2\}=-{\textstyle{\frac{i}{\sqrt{2}}}}g, \quad
\{{\bar G}_1,{\bar S}_2\}=-2iJ_{-},\quad \{{\bar G}_2,S_1\}={\textstyle{\frac{i}{\sqrt{2}}}}g,
\nonumber\\[2pt]
&&
\{{\bar G}_2,S_2\}=-2iD-2J_3, \quad
\{{\bar G}_2,{\bar S}_1\}=2iJ_{-},\quad \{S_1,{\bar S}_1\}=-2iK, \quad \{S_2,{\bar S}_2\}=-2iK,
\nonumber\\[2pt]
&&
\{S_1,J_{-}\}={\bar S}_2,\quad
\{S_1,J_3\}={\textstyle{\frac{i}{2}}} S_1, \quad \{S_2,J_{-}\}=-{\bar S}_1,\quad
\{S_2,J_3\}={\textstyle{\frac{i}{2}}} S_2,\quad \{{\bar S}_1,J_{+}\}=S_2,
\nonumber\\[2pt]
&&
\{{\bar S}_1,J_3\}=-{\textstyle{\frac{i}{2}}} {\bar S}_1, \quad
\{{\bar S}_2,J_{+}\}=-S_1,\quad
\{{\bar S}_2,J_3\}=-{\textstyle{\frac{i}{2}}} {\bar S}_2, \quad \{D,S_1\}={\textstyle{\frac{1}{2}}} S_1,
\nonumber\\[2pt]
&&
\{D,S_2\}={\textstyle{\frac{1}{2}}} S_2,\quad
\{D,{\bar S}_1\}={\textstyle{\frac{1}{2}}} {\bar S}_1, \quad \{D,{\bar S}_2\}={\textstyle{\frac{1}{2}}}
{\bar S}_2, \quad \{H,S_1\}=G_1,
\quad \{H,S_2\}=G_2,
\nonumber\\[2pt]
&&
\{H,{\bar S}_1\}={\bar G}_1, \quad \{H,{\bar S}_2\}={\bar G}_2,
\quad
\{H,D \}=H, \quad \{H,K \}=2D, \quad \{D,K \}=K,
\nonumber\\[2pt]
&&
\{J_{+},J_{-}\}=-2i J_3, \quad \{J_{+},J_3\}=iJ_{+}, \quad \{J_{-},J_3\}=-iJ_{-}.
\eea

\end{document}